\documentclass[twocolumn,amsmath,nofootinbib]{revtex4} 
\usepackage{graphicx}
\usepackage{url}
\begin{document}
\input{epsf.sty}

\def\affilmrk#1{$^{#1}$}
\def\affilmk#1#2{$^{#1}$#2;}

\def\affilmrk#1{$^{#1}$}
\def\affilmk#1#2{$^{#1}$#2;}
\def\be{\begin{equation}}
\def\ee{\end{equation}}
\def\bea{\begin{eqnarray}}
\def\eea{\end{eqnarray}}

\title{Observing Dark Energy Dynamics with Supernova,
 Microwave Background and Galaxy Clustering }

\author{Jun-Qing Xia$^1$, Gong-Bo Zhao$^{1}$, Bo Feng$^{2}$, Hong Li$^{1}$
and
Xinmin
Zhang$^{1}$ }

\affiliation{${}^1$Institute of High Energy Physics, Chinese Academy
of Science, P.O. Box 918-4, Beijing 100049, P. R. China}

\affiliation{ ${}^2$ Research Center for the Early Universe(RESCEU),
Graduate School of Science, The University of Tokyo, Tokyo 113-0033,
Japan}

\begin{abstract}

Observing dark energy dynamics is the most important aspect of the
current dark energy research. In this paper we perform a global
analysis of the constraints on the property of dark energy from the
current astronomical observations. We pay particular attention to
the effects of dark energy perturbations. Using the data from SNIa
(157 "gold" sample), WMAP and SDSS we find that the best fitting
dark energy model is given by the dynamical model with the equation
of state across -1. Nevertheless the standard $\Lambda$CDM models
are still a very good fit to the current data and evidence for
dynamics is currently not very strong. We also consider the
constraints with the recent released SNIa data from SNLS.

\end{abstract}


\maketitle


\setcounter{footnote}{0}

\section{Introduction}

In 1998 the analysis of the redshift-distance relation of Type Ia
supernova (SNIa) has established that our universe is currently
accelerating~\cite{Riess98,Perl99}. Recent observations of SNIa have
confirmed the accelerated expansion at high confidence
level\cite{Tonry03,Riess04,Riess05}. The nature of dark energy(DE),
the mysterious power to drive the expansion, is among the biggest
problems in modern physics and has been studied widely. A
cosmological constant, the simplest candidate of DE where the
equation of state (EOS) $w$ remains -1, suffers from the well-known
fine-tuning and coincidence problems\cite{SW89,ZWS99}.
Alternatively, dynamical dark energy models with the rolling scalar
fields have been proposed, such as quintessence\cite{pquint,quint},
the ghost field of phantom\cite{phantom} and the model of k-essence
which has non-canonical kinetic term\cite{Chiba:1999ka,kessence}.

Given that currently we know very little on the theoretical aspects
of dark energy, the cosmological observations play a crucial role in
our understanding of dark energy. The model of phantom has been
proposed in history due to the mild preference for a constant EOS
smaller than $-1$ by the observations\cite{phantom}. Although in
this scenario dark energy violates the weak energy condition(WEC)
and leads to the problem of quantum instabilities\cite{Phtproblms},
it remains possible in the description of the nature of dark
energy\cite{SPhtproblms}.

An intriguing aspect in the study of dark energy is that the recent
SNIa observations from the HST/GOODS program\cite{Riess04}, combined
together with the previous supernova data, somewhat favor the
dynamical dark energy model with an equation of state getting across
-1 during the evolution of the
universe\cite{sahni,Nesseris:2004wj,cooray,quintom}. Although the
conventional scalar dark energy models also show dynamical behaviors
with redshift, due to the instabilities of perturbations they cannot
preserve the required behavior of crossing the cosmological constant
boundary\cite{Vikman,Zhao:2005vj,Abramo:2005be}. The required model
of dark energy has been called as quintom\cite{quintom} in the sense
that its behavior resemble the combined behavior of quintessence and
phantom. However, the quintom models can be very different from the
quintessence or phantom in the determination of the evolution and
fate of the universe\cite{Feng:2004ff}. There are a lot of interests
in the literature recently in the building of quintom-like models.
With minimally coupled to gravity a simple realization of quintom
scenario is a model with the double fields of quintessence and
phantom\cite{quintom,guozk,hu,zhang,michael,xfzhang,Wei:2005nw}. In
such cases quintom would typically encounter the problem of quantum
instability inherited in the phantom component.
 However in the case of the single field scalar model of
quintom, Ref.\cite{li} added a high derivative term to the kinetic
energy and its energy-momentum tensor is equivalent to the two-field
quintom model. Such a model with a high derivative term is possibly
without quantum instabilities and as indicated by the SNIa
observations, we are living with the ghosts
\cite{hawking}\footnote{For another interesting single field quintom
model building see \cite{Huang:2005gu}.}.
 Perturbations of the quintom-like models have been studied
extensively in Ref.\cite{Zhao:2005vj}.

Observing the dark energy dynamics is currently the most important
aspect of the dark energy study. Besides the SNIa data, a thorough
investigation demands a fully consistent analysis of Cosmic
Microwave Background(CMB), large scale structure(LSS) and  with
multi-parameter freedoms. The aim of current paper is to study the
observational implications on dark energy in the consistent way. We
extend our previous work of Ref.\cite{Zhao:2005vj} and study the
full observational constraints on dynamical dark energy. In
particular we pay great attention to the effects of dark energy
perturbations when the equation of state gets across -1. Our paper
is structured as follows: in Section II we describe the method and
the data; in Section III we present our results on the determination
of cosmological parameters with the first year Wilkinson Microwave
Anisotropy Probe (WMAP) \cite{wmap}, SNIa \cite{Riess04,snls} and
Sloan Digital Sky Survey (SDSS) \cite{sloan} data by global fittings
using the Markov chain Monte Carlo (MCMC) techniques~
\cite{MCMC97,MacKayBook,Neil93}; finally we present our conclusions
in Section IV.

\section{Method and data}
In this section we firstly present the general formula of the dark
energy perturbations in the full parameter space of $w$ especially
when it crosses over the cosmological constant boundary. In our MCMC
fittings to WMAP, SNIa and SDSS, we adopt a specific parametrization
of the equation of state.

Despite our ignorance of the nature of dark energy, it is natural
to consider the DE fluctuation whether DE is regarded as scalar
field or fluid. In the extant cases like the two-field-quintom
model as well as the single field case with a high derivative
term~\cite{Zhao:2005vj}, the perturbation of DE is shown to be
continuous when the EOS gets across -1. For the conventional
parameterized equation of state one can easily reconstruct the
potential of the scalar dark energy if the EOS does not get across
-1. Resembling the multi-field model of quintom or quintom with
high derivative terms, the potential of the quintom dark energy
can be directly reconstructed from the parameterized EOS on either
side of the cosmological constant boundary\footnote{Although the
multi-field dark energy models are more challenging on theoretical
aspects of naturalness, given that we know very little on the
nature of dark energy, the energy momentum of such models can be
identified with single field scalar dark energy with high
derivative kinetic terms\cite{li}. Our phenomenological formula of
perturbations on DE corresponds to such models of multi-field
(quintom) with a negligible difference around the crossing point
of -1\cite{Zhao:2005vj}. }. In this paper we give a
self-consistent method to handle the perturbation in all the
allowed range of EOS especially for the region where EOS evolves
close to and crosses -1.

For the parametrization of the EOS which gets across -1, firstly we
introduce a small positive constant $\epsilon$ to divide the full
range of the allowed value of the EOS $w$ into three parts: 1) $ w
> -1 + \epsilon$; 2) $-1 + \epsilon \geq w  \geq-1 - \epsilon$; and 3) $w < -1 -\epsilon $.
Working in the conformal Newtonian gauge, one can describe the
perturbations of dark energy as follows \cite{ma}: \bea
    \dot\delta&=&-(1+w)(\theta-3\dot{\Phi})
    -3\mathcal{H}(c_{s}^2-w)\delta~~, \label{dotdelta}\\
\dot\theta&=&-\mathcal{H}(1-3w)\theta-\frac{\dot{w}}{1+w}\theta
    +k^{2}(\frac{c_{s}^2\delta}{{1+w}}+ \Psi)~~ . \label{dottheta}
\eea

Neglecting  the entropy perturbation contributions, for the regions
1) and 3) the EOS is always greater than $-1$ and less than $-1$
respectively and perturbation is well defined by solving
Eqs.(\ref{dotdelta},\ref{dottheta}). For the case 2), the
perturbation of energy density $\delta$ and divergence of velocity,
$\theta$, and the derivatives of $\delta$ and $\theta$ are finite
and continuous for the realistic quintom dark energy models. However
for the perturbations of the parametrizations there is clearly a
divergence. In our study for such a regime, we match the
perturbation in region 2) to the regions 1) and 3) at the boundary
and set\cite{Zhao:2005vj}
\begin{equation}\label{dotx}
  \dot{\delta}=0 ~~,~~\dot{\theta}=0 .
\end{equation}
In our numerical calculations we've limited the range to be $|\Delta
w = \epsilon |<10^{-5}$ and we find our method is a very good
approximation to the multi-field quintom, with the accuracy being
greater than $99.999\%$.

In the study of this paper the parameterized EOS of dark energy is
taken by\cite{Linder:2002et}
\begin{equation}\label{linder}
    w(z)=w_{0}+w_{1}\frac{z}{1+z}  .
\end{equation}
The method we adopt is based on the publicly available Markov Chain
Monte Carlo package \texttt{cosmomc}\cite{Lewis:2002ah,IEMCMC},
which has been modified to allow for the inclusion of dark energy
perturbations with EOS getting across -1\cite{Zhao:2005vj}. We
sample the following 8 dimensional set of cosmological parameters:
\begin{equation}\label{para}
    \textbf{p}\equiv(\omega_{b},\omega_{c},\Theta_S,\tau,w_{0},w_{1},n_{s},\log[10^{10} A_{s}])
\end{equation}
where $\omega_{b}=\Omega_{b}h^{2}$ and $\omega_{c}=\Omega_{c}h^{2}$
are the physical baryon and cold dark matter densities relative to
critical density, $\Theta_S$ is the ratio (multiplied by 100) of the
sound horizon and angular diameter distance, $\tau$ is the optical
depth, $A_{s}$ is defined as the amplitude of initial power spectrum
and $n_{s}$ measures the spectral index. Basing on the Bayesian
analysis, we vary the above 8 parameters fitting to the
observational data with the MCMC method. Throughout, we assume a
flat universe and
take the weak priors as: 
$\tau<0.8, 0.5<n_{s}<1.5, -3<w_{0}<3, -5<w_{1}<5$, a cosmic age
tophat prior as 10 Gyr$<t_{0}<$20 Gyr. Furthermore, we make use of
the HST measurement of the Hubble parameter $H_0 = 100h \quad
\text{km s}^{-1} \text{Mpc}^{-1}$ \cite{freedman} by multiplying the
likelihood by a Gaussian likelihood function centered around
$h=0.72$ and with a standard deviation $\sigma = 0.08$. We impose a
weak Gaussian prior on the baryon and density $\Omega_b h^2 = 0.022
\pm 0.002$ (1 $\sigma$) from Big Bang nucleosynthesis\cite{bbn}.

In our calculations we have taken the total likelihood to be the
products of the separate likelihoods of CMB, SNIa and LSS.
Alternatively defining $\chi^2 = -2 \log {\bf \cal{L}}$, we get
\be \chi^2_{total} = \chi^2_{CMB}+ \chi^2_{SNIa}+\chi^2_{LSS}~~~~
.\ee In the computation of CMB we have included the first-year
temperature and polarization data \cite{wmap,hinshaw} with the
routine for computing the likelihood supplied by the WMAP team
\cite{Verde03}. In the computation of nonlinear evolution of the
matter power spectra we have used the code of HALOFIT
\cite{halofit} and fitted to the 3D power spectrum of galaxies
from the SDSS\cite{sloan} using the code developed in Ref.
\cite{sdssfit}. In the calculation of the likelihood from SNIa we
have marginalized over the nuisance parameter. For the main
results of the current paper the supernova data we use are the
"gold" set of 157 SNIa published by Riess $et$ $al$ in
\cite{Riess04}. In addition, we also consider the constraints from
the distance measurements of the 71 high redshift type Ia
supernova discovered during the first year of the 5-year Supernova
Legacy Survey (SNLS)\cite{snls}.

\section{Results }

In this section we present our results, particularly focusing on the
effects of the dark energy perturbation. we start with the
descriptions on the background parameters, then present the
constraints on dark energy parameters. At last we give our
constraints on dark energy from the recent observational data of
SNLS.

\begin{table*}
\noindent 
 TABLE 1.  Mean $1\sigma$ constrains on
cosmological parameters using different combination of WMAP, SNIa
and SDSS information with/without DE perturbation. For the weakly
constrained parameters we quote the $95\%$ upper limit instead.
\begin{center}

\begin{tabular}{|c|ccc||ccc|}
  \hline

                                   &\multicolumn{3}{c||}{With DE Perturbation}                   &\multicolumn{3}{c|}{Without DE
                                   Perturbation} \\ 

  &\multicolumn{1}{c}{WMAP}&\multicolumn{1}{c}{WMAP+SN}&\multicolumn{1}{c||}{WMAP+SN+SDSS}
        &\multicolumn{1}{c}{WMAP}
        &\multicolumn{1}{c}{WMAP+SN}&\multicolumn{1}{c|}{WMAP+SN+SDSS}\\

  \hline\
  $\Omega_b h^2$     &$0.0232^{+0.0010}_{-0.0011}$    &$0.0234^{+0.0010}_{-0.0011}$    &$0.0232\pm0.0010$

                     &$0.0235\pm0.0013$  &$0.0232\pm0.0011 $&
$0.0230\pm0.0009$\\

  $\Omega_c h^2$     &$0.124\pm0.016     $      &$0.128\pm0.018$       &$0.123\pm0.010$

                     &$0.111\pm0.020     $      &$0.119\pm0.018$       &$0.122\pm0.010$  \\

 $\Theta_S$           &$1.046\pm0.006    $     &$1.047\pm0.006$
 &$1.046\pm0.005$

                     &$1.046\pm0.006    $     &$1.046\pm0.006$       &$1.046\pm0.005$  \\

  $\tau$             &$<0.256(95\%)     $     &$<0.264(95\%)$        &$<0.256(95\%)$

                     &$<0.399(95\%)     $     &$<0.324(95\%)$        &$<0.246(95\%)$   \\

  $w_0$              &$-0.732^{+0.623}_{-0.613}$  & $-1.172^{+0.231}_{-0.226}$  &$-1.167^{+0.191}_{-0.190}$

                     &$-0.617^{+0.193}_{-0.190}$  & $-1.080^{+0.105}_{-0.087}$  &$-1.098^{+0.078}_{-0.080}$   \\

  $w_1$               &$<1.59(95\%)     $     &$0.361^{+0.842}_{-0.883}$        &$0.597^{+0.657}_{-0.713}$

                      &$<0.832(95\%)     $     &$0.359^{+0.287}_{-0.179}$        &$0.416^{+0.293}_{-0.153}$   \\

  $n_s$               &$0.977^{+0.029}_{-0.030}     $    &$0.986^{+0.030}_{-0.031}$        &$0.982\pm0.030$

                      &$0.995^{+0.045}_{-0.041}     $    &$0.981^{+0.032}_{-0.033}$        &$0.970^{+0.024}_{-0.025}$   \\

  $\log[10^{10} A_s]$  &$3.181\pm0.134     $     &$3.207\pm0.132$        &$3.180^{+0.125}_{-0.123}$

                      &$3.245^{+0.202}_{-0.183}    $     &$3.206^{+0.153}_{-0.150}$        &$3.157^{+0.118}_{-0.119}$   \\
  \hline\
  $\Omega_\Lambda$    &$0.703^{+0.073}_{-0.072}     $    &$0.678\pm0.045$         &$0.681^{+0.031}_{-0.030}$

                      &$0.681\pm0.086     $    &$0.697\pm0.048$               &$0.686\pm0.031$\\

  $Age/GYr$           &$13.45\pm0.27     $     &$13.57^{+0.30}_{-0.29}$         &$13.67\pm0.24$

                      &$13.60\pm0.30     $     &$13.62\pm0.29$         &$13.67\pm0.23$  \\

  $\Omega_m$          &$0.297^{+0.072}_{-0.073}     $     &$0.322\pm0.045$         &$0.319^{+0.030}_{-0.031}$

                      &$0.319\pm0.086     $     &$0.303\pm0.048$         &$0.314\pm0.031$  \\

  $\sigma_8$          &$0.927^{+0.152}_{-0.154}     $     &$0.913^{+0.148}_{-0.150}$         &$0.854^{+0.096}_{-0.097}$

                      &$0.818\pm0.120     $     &$0.890^{+0.117}_{-0.118}$         &$0.882^{+0.073}_{-0.072}$   \\

  $z_{re}$            &$14.35^{+4.71}_{-4.68}     $     &$14.72^{+4.90}_{-4.77}$       &$14.41^{+4.97}_{-4.83}$

                      &$17.17^{+6.94}_{-6.27}     $     &$15.49^{+5.69}_{-5.38}$       &$13.73^{+4.78}_{-4.72}$   \\

  $H_0$               &$71.71^{+7.95}_{-8.02}     $     &$68.66^{+2.44}_{-2.45}$         &$67.90^{+2.48}_{-2.46}$

                      &$66.04^{+6.41}_{-6.47}     $     &$68.89^{+2.52}_{-2.61}$         &$68.07^{+2.35}_{-2.33}$   \\
\hline

  $\chi^{2}/d.o.f$    & 1429.1/1342 &1610.4/1499  &1633.2/1518
                      & 1428.4/1342&1610.6/1499 & 1634.1/1518\\

  \hline
\end{tabular}
\end{center}
\end{table*}

In Table 1 we list the mean 1$\sigma$ constraints on the
parameters with and without DE perturbations. We find that almost
all the cosmological parameters are well determined in our both
cases. However the reionization depth seems to be an exception
where a vanishing $\tau$ cannot be ruled out. We notice the
determination on $\tau$ is prior dependent. For example the WMAP
collaboration have taken a prior like
$\tau<0.3$\cite{Verde03,spergel03,peiris03}, which leads to a
relatively stringent constraint on $\tau$ by the observations and
a nonzero $\tau$ is particularly favored by the high power of
temperature-polarization cross correlation on the largest
scales\cite{kogut03}. However when the strong prior on $\tau$ is
dropped one will in general get a less stringent bound from the
full observational constraints, as also shown in Ref.
\cite{sdssfit}. The prior on $\tau$ is somewhat crucial for our
parameter estimation since its effects on CMB can be compensated
with the tilt of the primordial scalar as well as the tensor
spectrum. As we will show below it will also be correlated with
the dark energy parameters due to the Integrated Sachs-Wolfe(ISW)
effects.

\begin{figure}[htbp]
\begin{center}
\includegraphics[scale=0.4]{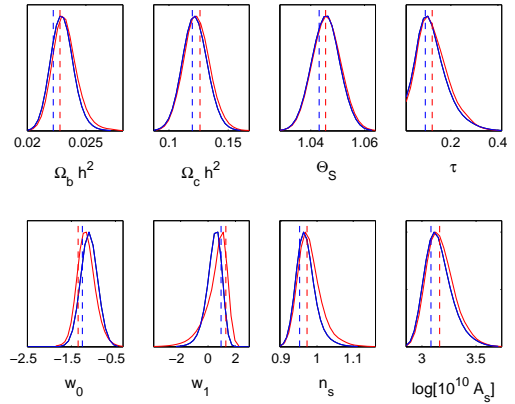}
\vskip-1.3cm \vspace{10mm}\caption{1-D constrains on individual
parameters using WMAP+157 "gold" SNIa+SDSS with our 8-parameter
parametrization discussed in the text. Solid curves illustrate the
marginalized distribution of each parameter with(red) and
without(blue) DE perturbation. Dotted vertical lines shows the
quantity of every parameter with (red) and without (blue) DE
perturbation giving the maximum likelihood. \label{1d}}
\end{center}
\end{figure}

\begin{figure}[htbp]
\begin{center}
\includegraphics[scale=0.38]{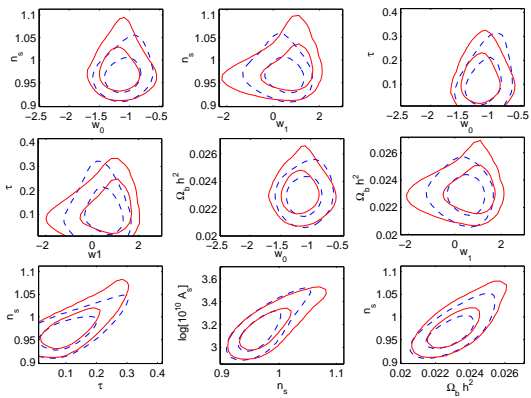}
\vskip-1.3cm \vspace{10mm}\caption{Sampled 2-D contours of the
background parameters and also the contours among dark energy and
the background parameters. Here we use the same parametrization and
data sets as FIG.2, red solid and blue dashed lines are for
perturbated and unperturbated DE respectively. \label{2d}}
\end{center}
\end{figure}

In Fig.\ref{1d} also we delineate the corresponding posterior one
dimensional marginalized distributions of the cosmological
parameters from the combined observations of WMAP, SDSS and SNIa.
The dotted vertical lines shows the quantity of every parameter with
(red) and without (blue) DE perturbation giving the maximum
likelihood.  Due to the fact that the peaks in the likelihood are
different from the corresponding expectation values, the dashed
lines in Fig.\ref{1d} do not lie at the center of the projected
likelihoods. A vanishing $\tau$ cannot be excluded by the full
combined observations at high confidence level. For the order of
parameters listed in (5)
 the best fit values constrained by the full
dataset (WMAP + SNIa + SDSS) is
\textbf{p}=(0.023,0.12,1.04,0.16,-1.30,1.25,0.995,3.23). And for
comparison the resulting parameters when switching off dark energy
perturbations are given by
\textbf{p}=(0.023,0.12,1.05,0.14,-1.15,0.63,0.962,3.14). Comparing
with the bottom of Table 1, although the minimum $\chi ^2$ values
have not been affected significantly (up to 1) by dark energy
perturbations, all the best fit parameters have been changed.
Moreover, the allowed parameter space has been changed a lot and
the constraints on the background parameters have been less
stringent when including the dark energy perturbations. This can
also be clearly seen from the two dimensional contour plots on the
background parameters in Fig.\ref{2d}. The reason is not difficult
to explain. The ISW effects of the dynamical dark energy boosts
the large scale power spectrum of CMB\cite{Zhao:2005vj}. For a
constant equation of state Ref. \cite{WL03} has shown that when
the perturbations of dark energy have been neglected incorrectly,
a suppressed ISW will be resulted for quintessence-like dark
energy and on the contrary, an enhanced ISW is led to by
phantom-like dark energy. In this sense if we neglect dark energy
contributions, there will be less degeneracy in the determination
of dark energy as well as the relative cosmological parameters.
However, dark energy perturbations are anti-correlated with the
source of matter perturbations and this will lead to a
compensation on the ISW effects, which result in a large parameter
degeneracy\cite{WL03}. In fact as we have shown that crossing over
the cosmological constant boundary would not lead to distinctive
effects\cite{Zhao:2005vj}, hence the effects of our smooth
parametrization of EOS on CMB can also be somewhat identified with
a constant effective equation of state\cite{Wang:1999fa}
\begin{equation}\label{weff}
    w_{eff}\equiv\frac{\int da \Omega(a) w(a)}{\int da \Omega(a)}~~,
\end{equation}
however the SNIa and LSS observations will break such a degeneracy.
Thus for the realistic cases of including dark energy perturbations,
the correlations between the dark energy and the background
parameters as well as the auto correlations of the background
cosmological parameters have been enlarged, as can be seen from
Fig.\ref{2d}.

\begin{figure}[htbp]
\begin{center}
\includegraphics[scale=0.55]{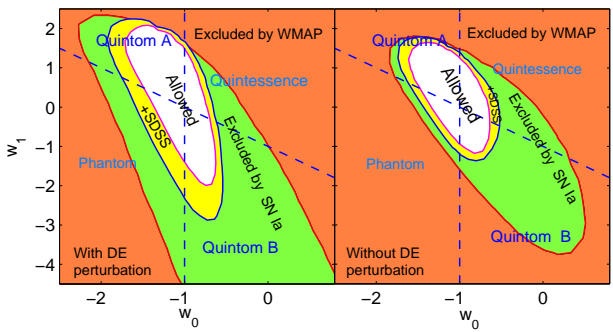}
\vskip-1.3cm \vspace{10mm}\caption{95\% constrains in the
($w_{0},w_{1}$) plane with and without dark energy perturbation
from left to right. Shaded dark orange region is excluded by WMAP
only for our 8-parameter estimation. The dashed lines stand for
$w_0=-1$ and $w_0 + w_1 = -1$, see the text for details.
\label{compare}}
\end{center}
\end{figure}

The contribution of dark energy perturbation affect significantly
the distribution of  $w_0$ and $w_1$, which can also be seen from
Fig.\ref{compare} on the constrains in the $(w_0, ~ w_1 )$ plane.
For the parameters $( w_0 , ~ w_1)$ the inclusion of the dark
energy perturbation change its best fit values from $(-1.15,
~~0.63)$ to $(-1.30, ~~1.25)$. In Fig.\ref{compare}, from outside
in, the contours shrink with adding 157 SNIa data provided by
Riess $et$ $al$ and SDSS information. Dark energy perturbation
introduces more degeneracy between $w_{0}$ and $w_{1}$ thus
enlarges the contours significantly. For the discussion on
dynamical dark energy we have separated the space of $w_0 - w_1$
into four areas by the lines of $w_0=-1$ and $w_0 + w_1 = -1$. The
areas will then represent the quintessence where the EOS is always
no less than -1 (the area with $w > 1$ can be reached by
quintessence with a negative potential\cite{Felder:2002jk}),
Quintom A where $w$ is phantom-like today but quintessence-like in
the past, phantom where the EOS is always no larger than -1 and
Quintom B where dark energy has $w>-1$ today but $w<-1$ at higher
redshifts. From the figure we can see that dynamical dark energy
with the four types are all allowed by the current observations
and Quintom A seems to cover the largest area in the 2-dimensional
contours with all the data we used.

\begin{figure}[htbp]
\begin{center}
\includegraphics[scale=0.8]{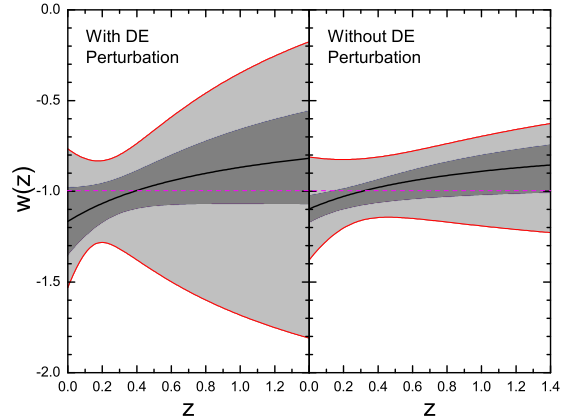}
\vskip-1.3cm \vspace{10mm}\caption{Constrains on w(z) using WMAP +
157 "gold" SNIa data + SDSS with/without DE perturbation.
Median(central line), 68\%(inner, dark grey) and 95\%(outer, light
grey) intervals of w(z) using 2 parameter expansion of the EOS in
(4).  \label{w}}
\end{center}
\end{figure}

As shown in Fig.\ref{compare} that $w_{0}$ and $w_{1}$ are in
strong correlations. The constraints on $w(z)$ are perhaps
relatively model independent, as suggested by Ref.\cite{seljak04}.
Following\cite{HT} we obtain the constraints on $w(z)$ by
computing the median and 1, 2$\sigma$ intervals at any redshift.
In Fig.\ref{w} we plot the behavior of the dark energy EOS as a
function of redshift $z$, we find that at redshift $z=0.3$ the
constraint on the EOS is relatively the most stringent. One can
see that the perturbation reinforces the trend of DE to cross -1
at $z\sim 0.3$. However due to the limitation of the observational
data, the quintom scenario is only favored at $1 \sigma$ by the
full dataset of WMAP, SDSS and SNIa. We find the value at $z=0.3$
is restricted at \be
 w(z=0.3)={-1.002^{+0.044+0.180}_{-0.079-0.159}}
\ee
 for the case without dark
energy perturbations and \be
 w(z=0.3)={-1.029 ^{+0.108+0.230}_{-0.098-0.288}}
\ee when including dark energy perturbations.  Correspondingly at
redshift $z=1$ the constraints turn out to be \be
 w(z=1)={-0.890^{+0.180+0.193}_{-0.159-0.301}}
\ee without perturbations and \be
 w(z=1)={-0.868^{+0.215+0.520}_{-0.204-0.815}}
\ee when including dark energy perturbations. One should bear in
mind that such a constraint is not really model independent, as
shown in Refs. \cite{BCK,xia}.

\begin{figure}[htbp]
\begin{center}
\includegraphics[scale=0.33]{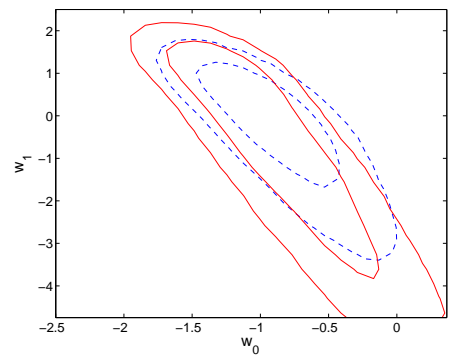}
\vskip-1.3cm \vspace{10mm}\caption{Two dimensional constraints on
the parameters of dynamical dark energy from the combined
constraints of SNLS, WMAP and SDSS. The solid and dotted lines are
the constraints with and without dark energy perturbations,
respectively.\label{snls}}
\end{center}
\end{figure}

Recently the authors of Ref.\cite{snls} made the distance
measurements to 71 high redshift type Ia supernovae discovered
during the first year of the 5-year Supernova Legacy Survey
(SNLS). SNLS will hopefully discover around 700 type Ia
supernovae, which is an intriguing ongoing project. Following
Ref.\cite{snls} we combine the "new 71 high redshift SNIa data
$\bigoplus$ the 44 nearby SNIa", together with WMAP and SDSS.  We
plot the constrains on the dark energy parameters in
Fig.{\ref{snls}}. We find the current data of SNLS are very weak
in the determination of dark energy parameters. Although our best
fit values are given with a quintom like dark energy: $( w_0 , ~
w_1)= (-1.47,1.44)$, a cosmological constant fits well with SNLS
in the 1$\sigma$ region.\footnote{Ref. \cite{NP05} made some study
on SNLS implications of dynamical dark energy using SNIa data
only.}

\section{discussion and conclusion}

In this paper we have performed an analysis of global fitting on the
EOS of dark energy from the current data of SNIa, WMAP and SDSS.
Dark energy perturbation leaves imprints on CMB through ISW effects
and changes the matter power spectrum by modifying the linear growth
factor as well as the transfer function. Our results show that when
we include the perturbations of dark energy, the current
observations allow for a large variation in the EOS of dark energy
with respect to redshift\footnote{Ref. \cite{tao05} studied the
perturbations of dynamical dark energy only for the regime where
$w>-1$ and Ref. \cite{ex} considered both the cases for $w>-1$ and
$w<-1$, but did not include the perturbations for quintom like dark
energy; previous global analysis like Ref. \cite{seljak04,ex1} did
not consider dark energy perturbations.}. A dynamical dark energy
with the EOS getting across $-1$ is favored at 1$\sigma$ with the
combined constraints from WMAP, SDSS and the "gold" dataset of SNIa
by Riess $et$ $al$. When we use the recently released SNLS data and
the nearby data of type Ia supernova instead, the parameter space is
enlarged and a cosmological constant is well within 1$\sigma$,
although a quintom dynamical dark energy is still mildly favored.

In our all results, the perturbation of dark energy plays a
significant role in the determination of cosmological parameters.
Neglecting the contributions of dark energy perturbation will lead
to biased results which are more stringent than the real cases. In
the next decade, there will be many ongoing  projects in the
precise determination of cosmological parameters. We can hopefully
detect the signatures of dynamical dark energy like quintom
through global fittings to the observations, where it is crucial
for us to include the contributions of dark energy perturbations.

{\bf Acknowledgements:} Our MCMC chains were finished in the
Shuguang 4000A system of the Shanghai Supercomputer Center(SSC).
This work is supported in part by National Natural Science
Foundation of China under Grant Nos. 90303004, 10533010 and
19925523 and by Ministry of Science and Technology of China under
Grant No. NKBRSF G19990754. We thank Mingzhe Li for helpful
discussions and Hiranya Peiris for comments on the manuscript.

\end{document}